\documentclass[10pt,journal,compsoc]{IEEEtran}
\usepackage{amsmath}
\usepackage{xcolor,soul,framed} 

\colorlet{shadecolor}{yellow}
\usepackage[pdftex]{graphicx}
\graphicspath{{../pdf/}{../jpeg/}}
\DeclareGraphicsExtensions{.pdf,.jpeg,.png}

\usepackage{array}
\usepackage{mdwmath}
\usepackage{mdwtab}
\usepackage{eqparbox}
\usepackage{url}
\usepackage{multirow}
\usepackage{tabularx}
\usepackage{tikz}
\usepackage{lipsum}
\ifCLASSOPTIONcompsoc
  \usepackage[nocompress]{cite}
\else
  \usepackage{cite}
\fi
\ifCLASSINFOpdf
\else
\fi
\begin{document}
%
\title{Modeling Effective Lifespan of Payment Channels}
%
%
%
%

\author{Soheil~Zibakhsh~Shabgahi,
      Seyed Mahdi~Hosseini, Seyed Pooya Shariatpanahi, Behnam Bahrak}
\IEEEtitleabstractindextext{%
\begin{abstract}
While being decentralized, secure, and reliable, Bitcoin and many other blockchain-based cryptocurrencies suffer from scalability issues. One of the promising proposals to address this problem is off-chain payment channels. Since, not all nodes are connected directly to each other, they can use a payment network to route their payments. Each node allocates a balance that is frozen during the channel's lifespan. Spending and receiving transactions will shift the balance to one side of the channel. A channel becomes unbalanced when there is not sufficient balance in one direction. In this case, we say the effective lifespan of the channel has ended.

In this paper, we develop a mathematical model to predict the expected effective lifespan of a channel based on the network's topology. We investigate the impact of channel unbalancing on the payment network and individual channels. We also discuss the effect of certain characteristics of payment channels on their lifespan. Our case study on a snapshot of the Lightning Network shows how the effective lifespan is distributed, and how it is correlated with other network characteristics. Our results show that central unbalanced channels have a drastic effect on the network performance. 
\end{abstract}

\begin{IEEEkeywords}
Bitcoin, Lightning Network, Payment Channel, Lifespan, Random Walk.
\end{IEEEkeywords}}

\maketitle

\IEEEdisplaynontitleabstractindextext

%
\IEEEpeerreviewmaketitle

\IEEEraisesectionheading{\section{Introduction}\label{sec:introduction}}

%
%
%
%

\IEEEPARstart{B}{itcoin} is the first decentralized cryptocurrency, introduced in 2008 which provides security, anonymity, transparency, and democracy without any trusted third party \cite{nakamoto2008Bitcoin}. Most of these properties are achieved by using a blockchain as a distributed ledger. An inherent problem with using a blockchain over a network is that it sacrifices scalability \cite{chauhan2018blockchain}, \cite{karame2016security}. The reason is that all nodes, potentially tens of thousands, must exchange, store, and verify each and every transaction in the system \cite{poon2016Bitcoin}. Furthermore, each block has a limited size and blocks get generated at regular intervals (approximately every 10 minutes). This means that with the current blocksize of 1\,MB the throughput of Bitcoin is about 4.6 transactions per second, which is much slower than centralized systems like Visa, WeChatPay, and PayPal \cite{avarikioti2020ride}; making the use of  Bitcoin in everyday transactions impractical.

Another trade-off the Bitcoin consensus makes is that it ensures security by waiting for other miners to confirm a transaction by extending the block holding that transaction, which reduces the throughput. This way it makes sure that the double spending attack is highly improbable. Currently, the standard waiting time for a block to be confirmed is 6 blocks, which is almost one hour \cite{karame2012double}. 

Bitcoin's capacity limitations are being felt by users in the form of increased transaction fees and latency. With an increasing demand for making transactions, users need to pay more transaction fees in order to make sure that their transaction is more profitable for the miners; hence have a higher chance of making it into a block. Queuing of transactions and network bandwidth will lead to a longer delay time for a transaction to appear in the blockchain.

There are many different proposals to solve the scalability problem.  Most of the proposals fall into three categories: $Layer0$, $Layer1$, and $Layer2$ solutions \cite{zhou2020solutions}. $Layer0$ solutions try to enhance the infrastructure, like the network that connects the nodes. $Layer1$ solutions try to enhance the blockchain's shortcomings by changing the consensus mechanism and protocols \cite{david2018ouroboros}, \cite{eyal2016bitcoin}. $Layer2$ solutions propose ways to move away from the blockchain, and for this reason, they are also called off-chain solutions \cite{poon2016bitcoin2}. 

In 2016 the idea of Lightning Network (LN) was proposed to move the transactions to the second layer (off-chain) \cite{poon2016Bitcoin}. 
The Lightning Network consists of payment channels in a P2P fashion. Payment channels allow two parties to exchange payments with negligible time and cost, but both parties must freeze an initial fund in the channel so no one can spend more money than they own and no double spending occurs. It is important to note that the sum of funds in each channel remains constant throughout the channel's lifespan and only the channel's balance changes.
When two parties that do not have a direct channel want to exchange payments they can use other parties to route their payments. So a network of nodes is constructed and all the connected nodes can send each other payments.

This system moves the cost of submitting a transaction off the blockchain. Only the final states between two nodes will eventually make it into the blockchain, which significantly increases throughput. Furthermore, no time is needed for the transaction to be confirmed and all transactions in a channel happen almost instantly.

After several transactions through a channel, the channel starts to get unbalanced; meaning all of its funds have gone to one of the parties and the other node cannot route any more payments through the channel. In this case, it is best to close the unbalanced channel or open a new one.

In this paper, we investigate the expected effective lifespan of a channel in a payment network. Our contributions can be summarized as follows:

\begin{itemize}

\item We provide simulation evidence of how channel unbalancing impacts its throughput. Moreover, we show how the performance of the payment network can be affected if a number of channels become unbalanced.

\item We present a mathematical model of payment channels to predict the expected time for a channel to get unbalanced considering the channel’s position in the network and its initial balance. We call this time the \emph{Effective Lifespan} of the channel.

\item We evaluate our model through simulation, and observe how the Effective Lifespan of a channel is affected if we change any of its characteristics.

\item By analyzing a recent snapshot of the Lightning Network, we find the distribution of real-world channel lifespans and its correlation with the network’s topological parameters. We also investigate the relationship between the centrality of a channel in the network and its effective lifespan.

\end{itemize}

\section{Related Work}
While the LN white paper \cite{poon2016Bitcoin} does not discuss channel re-balancing, there exists some research on channel balances and their significance.

The importance of channel balances is mainly discussed in four major areas; re-balancing, security, performance, and financial.

\textbf{Re-balancing}:
\cite{khalil2017revive} proposes a method for re-balancing payment channels. This work allows arbitrary sets of users to securely re-balance their channels. However, this paper does not discuss the application of re-balancing, and how frequently it should be performed. \cite{conoscenti2019hubs} also proposes methods for rebalancing LN channels, but does not discuss the frequency of rebalancing.

\textbf{Performance}:
In \cite{pickhardt2020imbalance} the authors discuss why it is in the best interest of the network to have balanced channels. They propose a method to re-balance some channels to improve the network’s performance. \cite{tikhomirov2020probing} presents a method in which a node can make its channels balanced through circular subgraphs. It also develops a method for measuring imbalance in a payment network.

\textbf{Security}:
There has been some research on the security aspects of channel unbalancing. In \cite{tikhomirov2020probing}, \cite{herrera2019difficulty}, and \cite{kappos2021empirical} the authors describe a method in which it is possible for an adversary to uncover channel balances. Having unbalanced channels poses the threat of griefing attacks. The incentive for honest behavior in the LN channels is the penalty for misbehavior. If a node cheats by publishing an old contract, it will be penalized and all of the channel funds can be claimed by the victim. When channels are unbalanced the penalty is less so there is less incentive for honest behavior. In \cite{rahimpour2021hashcashed} the authors discuss some countermeasures like watchtowers to keep the misbehaving nodes from closing the channel.

\textbf{Financial}:
Routing payments through a channel can make revenue for the owner. So payment channels can be looked at as investments.
In \cite{branzei2017charge} the authors do an in-depth financial analysis on how much should payment channels charge for routing payments. One of the key factors in this analysis is the lifespan of payment channels. In order to analyze investing in a payment channel, nodes should be able to have an estimate on how long the investment stays profitable and what is the impact of channel unbalancing on the profits of a channel.  Branzei et al. \cite{branzei2017charge} assume an equal probability of having a payment from each side in a channel and use the lifetime of channels for financial analyses. We will show how the lifespan of a channel could be affected by this probability.

In this paper we focus on the details of estimating channel lifespans; considering parameters such as the placement of the channel in the topology and payment rates between each pair and explain the importance of estimating channel lifespans. This gives us a better and more realistic estimation of the channel's lifespan compared to existing work. Moreover, we measure the impact of imbalanced channels on the network.

Despite the importance of payment channel’s lifespan, to the best of our knowledge, the expected lifespan of channels in the payment network has not been discussed in detail.

\section{Technical Background}
In this section, we provide a technical background to understand the remainder of this paper thoroughly.

\subsection*{Payment Channels}
Payment channel is a financial contract between two parties in a cryptocurrency like Bitcoin. The contract allocates a balance of funds from both parties. The contract is established by a 2-of-2 multisignature address which requires the cooperation of both parties to spend the funds.

Payments are made off the blockchain by passing on a new version of the contract with a different balance of allocated funds on the spending transaction; which both parties have to sign. The channel is closed when one of the parties publishes the latest version of the contract to the blockchain. We define the payment direction to be the direction in which funds are moving during a transaction.

In this paper, we call the sum of locked funds in a channel the channel's capacity. 
When all of the funds of a channel are allocated to one of the parties, the channel becomes \emph{unbalanced}. In this case payments can only be made from one side of the channel. 
A channel’s \emph{effective lifespan} is the time from creation of a channel until the first imbalance occurs.
A channel's success probability is defined as the number of successful payments made through the channel divided by the total number of payment attempts.

Several connected payment channels can form a payment network, in the case of Bitcoin, this network is called the Lightning Network \cite{poon2016Bitcoin}. This network is used to route payments through intermediate channels between nodes who do not have a direct channel between them.
We define a network's success probability as the total number of successful payments made on the network divided by the total attempts to route  payments through the network \cite{conoscenti2018cloth}.

\subsection*{Random Walk}
The random walk model has been used in a wide variety of contexts to model the movement of objects in different spaces. This paper uses one-dimensional random walk to model the liquidity balance in a payment channel. Two endpoints on the left and right sides of the random walk are assumed to represent the channel imbalance condition.
In our model, each payment corresponds to one step of the random walk model, and the direction of the payment determines the direction of that step. Suppose we take probabilities $p$ and $1-p$ as the probability of payment direction (i.e., step direction). We can find the expected number of payments (steps) needed for the channel (the random walk model)to get unbalanced (to reach one of the endpoints).

\subsection*{Betweenness Centrality}
Betweenness centrality is a measure based on shortest paths for the importance of the location of a node or an edge in a graph. Betweenness centrality for an $edge(a,b)$ in the network is defined as follows: $\underset{s \neq t}{\sum^{}_{s,t \in V}} \frac{\sigma(s,t|edge(a,b))}{\sigma(s,t)}$, where $\sigma(s,t)$ is the total number of shortest paths between nodes $s$ and $t$ and $\sigma(s,t|edge(a,b))$ is the total number of shortest paths between $s$ and $t$ that pass through $edge(a,b)$.

\section{Motivation} \label{motivation}

\begin{figure}
  \begin{center}
  \includegraphics[width=3.0in]{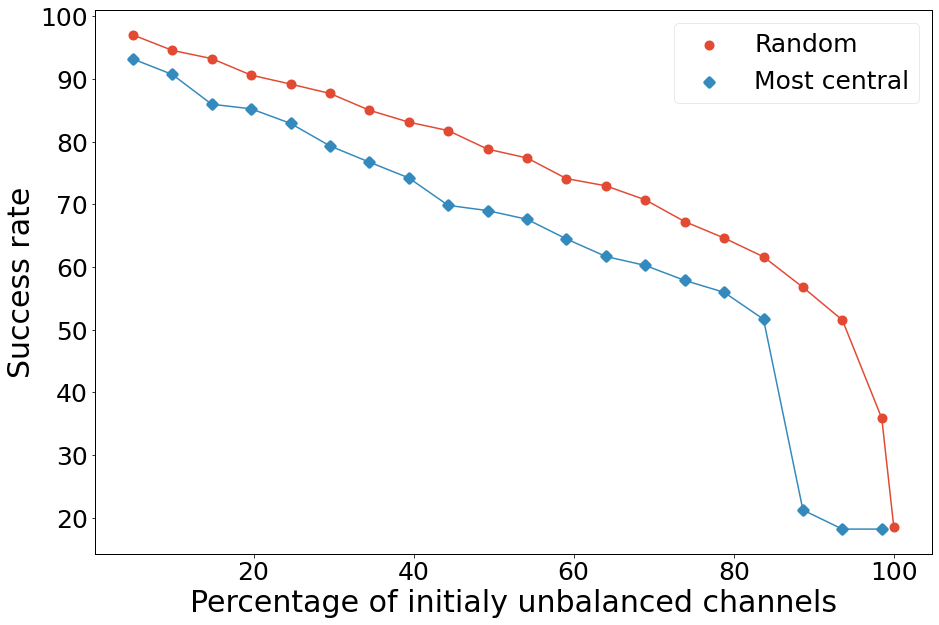}
  \caption{Relation of network success rate with percentage of unbalanced channels in the network.}
\label{unbalance_success}
  \end{center}
\end{figure}

One of the important characteristics of a payment network is reliability. Reliability can be defined as the probability of payment success \cite{conoscenti2018cloth}.

In this section, we analyze the payment routing failure probability of a singular channel after unbalancing, and the network's success probability of routing a payment when some channels are unbalanced.

\subsection{Singular Channel}
\begin{figure}
  \begin{center}
  \includegraphics[width=3.0in]{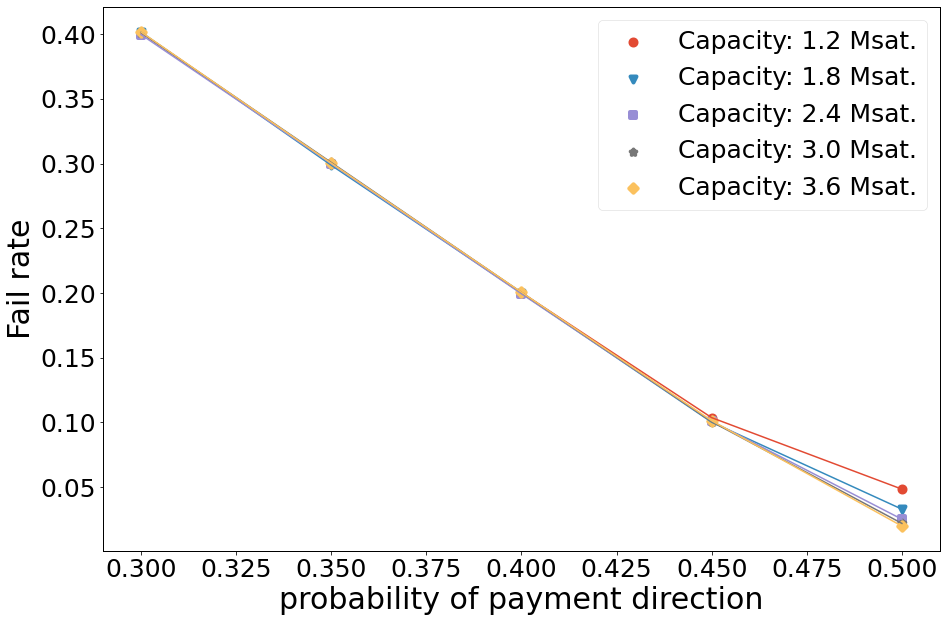}
  \caption{Failure rate after unbalancing.}
\label{unbalance_failure}
  \end{center}
\end{figure}

We ran a simulator of a single payment channel to see how much the failure rate increases after the first time that the channel becomes unbalanced. Fig.~\ref{unbalance_failure} shows the failure rate after the first time a channel becomes unbalanced. During the simulation, 5000 payments were being routed through an initially balanced channel. Then the simulator calculates the failure rate after the first time the channel becomes unbalanced. As Fig.~\ref{unbalance_failure} suggests, the probability of the direction of payments ($p$) is a key factor in determining how much the probability of payment success degrades after the first imbalance occurs. Channels capacity has little to no impact on how well it performs after unbalancing.

These results show that the probability of payment direction ($p$), which depends on the network topology and the network's transaction flow, is one of the most important parameters in determining the channel's lifespan; more importantly, shows the impact of unbalancing on channel success probability after the channel becomes unbalanced.

\subsection{Network Performance}
Using the CLoTH simulator \cite{conoscenti2018cloth} we simulate and measure the performance of Lightning Network. In each iteration we take channels from the given LN snapshot and make them unbalanced, we then measure the success probability after attempting $5000$ payments. Choosing more central channels as unbalanced channels is more reasonable, because they route more payments and thus have a higher probability of becoming unbalanced in the real world.
We considered two scenarios for selecting channels to unbalance: choosing channels randomly and choosing channels that have a higher betweenness centrality. As illustrated in Fig. \ref{unbalance_success}, as the percentage of unbalanced channels increases, the routing success rate decreases dramatically for both channel selection scenarios. In the random selection scenario, it is noticeable that the first 10 percent of unbalanced channels have less effect on the network performance than the last 10 percent of unbalanced channels. We see that unbalancing channels with a higher betweenness centrality has a higher impact on the network performance in contrast with the random selection scenarios. Therefore, per any percentage of unbalanced channels, selection with betweenness centrality is more effective.

\begin{figure}
  \begin{center}
  \includegraphics[width=3.0in]{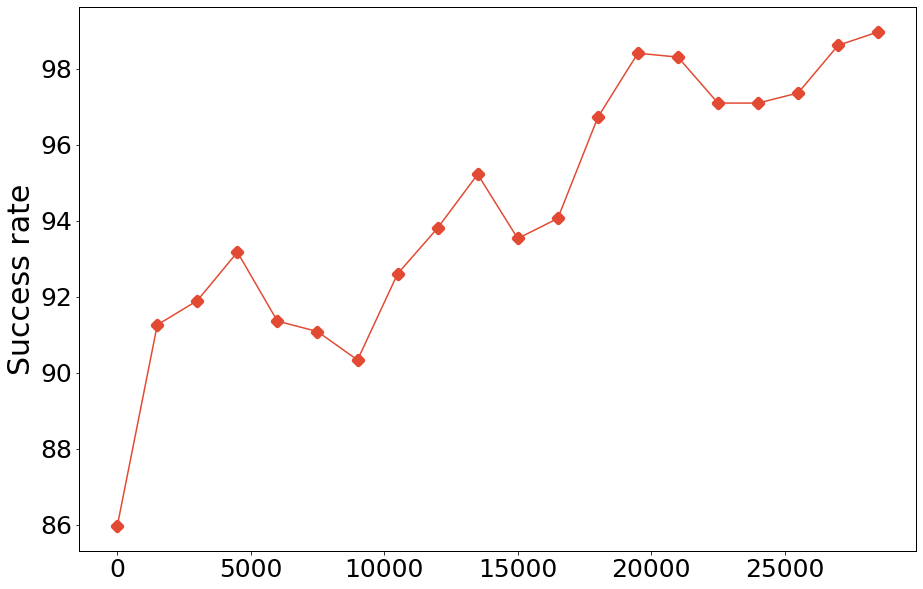}
  \caption{Per each data point the $i$-th to $(i+4500)$-th most central channels are unbalanced and the success rate of the network is measured. The total number of channels is 30457.}
\label{between_unbalance}
  \end{center}
\end{figure}

Seres et al. \cite{seres2020topological} suggest that in the Lightning Network, the top $14\%$ central channels will have the most significant impact on the network.
In a different experiment we made $15\%$ of the network’s channels unbalanced, we first sort the channels by betweenness centrality and take a window of $15\%$ of the channels per experiment. We start with the $15\%$ most central channels and move all the way up to $15\%$ least central channels. It can be inferred from the results in Fig. \ref{between_unbalance} that more central channels have more impact on the network success rate when they become unbalanced. As we can see in Fig. \ref{between_unbalance}, the top $15\%$ central channels have the most significant effect on the success rate when they become unbalanced. This confirms the result from Fig. \ref{unbalance_success}.

\section{The model}\label{Modeling}

As we discussed in Section~\ref{motivation} channel balances have a significant effect on both channel, and network performance. In this section, we introduce a mathematical model to determine the expected time for a channel to get unbalanced; we call this the channel's expected lifespan. We model the dynamics of a payment channel with a random walk problem. Each payment passing through the channel will represent a step the random walker takes. We will first discuss our assumptions and describe the model in detail. We then discuss how to find the model parameters. We proceed by doing an analysis on how the expected lifespan is affected by changing channel's characteristics.

\subsection {Random Walk Model} \label{Modeling:Random-Walk}

Take a payment channel between two nodes $A$ and $B$, and take their initial balance allocated for the channel to be $F_{A}$ and $F_{B}$, respectively. The goal is to determine the expected time it will take for this payment channel to become unbalanced for the first time. We make the following assumptions:

\begin{itemize}
  \item All the payments have the same amount denoted with $\omega$ (PaymentSize).
  \item The payments from each node come with a Poisson distribution. 
\end{itemize}

Since the number of nodes is large and the probability of sending a transaction for a given time is small, we can assume that transaction arrival for each channel is a Poisson process for moderate time windows \cite{hodges1960poisson}.
Although the dynamics of the network will change over time, we make the assumption of having a fixed topology.

We model the dynamics of a payment channel with a random walk problem. Each payment is simulated by a step the random walk takes.
To simulate a payment channel, take the liquidity of node $A$ as the distance of the random walk from the endpoint on the right hand side and the liquidity of node $B$ as the distance from the endpoint on the left hand side.

The payment direction determines the direction of that step. So the payment direction probability is the probability of going to the right or left for the random walk in each step.

Let the random walk start at the origin of the number line. The two endpoints $a$ and $-b$ are $\lfloor\frac{F_{A}}{\omega}\rfloor$ and $\lfloor\frac{F_{B}}{\omega}\rfloor$, respectively.

Since we assume that the payments from each side are made independently with a Poisson process, and the sum of two independent Poisson processes is itself a Poisson process, we can say that payments come to the channel with a Poisson distribution having:

\begin{equation} \label{sumRate}
    \lambda_{payment}=\lambda_{A,B}+\lambda_{B,A},
\end{equation}
thus the relation between expected time and expected number of random walk steps is: 
\begin{equation}\label{Expectedrateequation}
    E_{time} = \frac{E_{steps}}{\lambda_{payment}} ,
\end{equation}

where $E_{time}$ is the expected time until unbalancing and $E_{steps}$ is the expected number of steps until unbalancing occurs.

The expected number of payments until unbalancing occurs, can be a better metric depending on the application; when multiplied by average fee per payment, it gives the expected routing income, and when divided by $\lambda$ it gives the expected lifespan.

The objective is to determine the time it takes for a channel to become unbalanced. We first try to find the expected number of steps needed for the random walker to reach $+a$ or $-b$ for the first time.

\textbf{Lemma 1.} \textit{The expected number of steps to reach $+a$ or $-b$ for the first time starting from zero considering the probability $p$ for the positive direction and $q=1-p$ for the negative direction is: 
\begin{equation} \label{stepEquation}
  E_{steps} =
  \begin{cases}
    \frac{a p^a(p^b-q^b) + b q^b(q^a-p^a)}{(p-q)(p^{a+b} - q^{a+b})} & p \neq 1/2\\
      a b & p = 1/2
  \end{cases}
\end{equation}.}

We provide the proof of Lemma 1 in Appendix A.

\begin{figure}
  \begin{center}
  \includegraphics[width=3.0in]{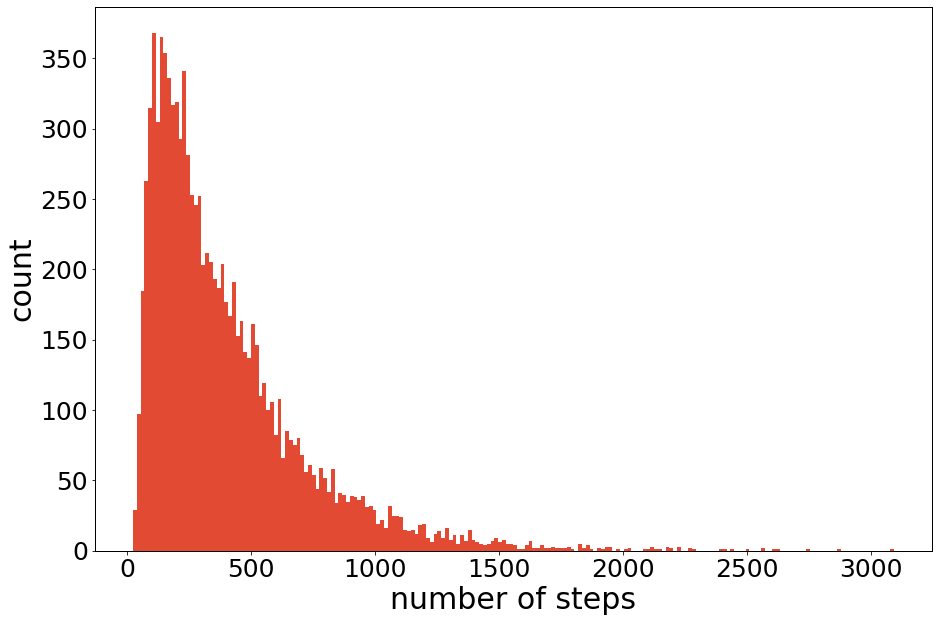}
  \caption{Distribution of expected lifespan with 10000 random walk simulations with $p=\frac{1}{2}$ and $a = b = 1.2\,Msat.$}
\label{time_hist}
  \end{center}
\end{figure}

We simulated a Random Walk which starts from point zero with the same probability of going to each side ($p=\frac{1}{2}$). The simulation ran 10000 times to find the distribution of the number of steps needed to reach $+a$ or $-b$. Fig. \ref{time_hist} illustrates the result of the simulation. We can observe that most of the times the random walk reaches one of the bounds in less than 400 steps, but there are not many situations where it takes a huge number of steps to reach these bounds. However, the average number of steps needed to reach these bounds is 400.5 confirming \ref{stepEquation}.

\subsection {Finding p}
In Section~\ref{Modeling:Random-Walk} we modeled the payment channel dynamics with a random walk and a parametric formula was constructed according to Lemma 1.

A payment network can be formally expressed by an unweighted directed graph. $V$ represents the set of all nodes, and the set of edges is denoted by $E$. Each channel is represented using two edges from $E$ each for one of the directions.

We define $MRates$ to be the matrix of payment rates between each two nodes. The rate of payments (i.e., number of payments per day) from node $i$ to node $j$ is denoted by $MRates_{ij}$.

$\lambda_{a,b}$ represents the rate of payments transmitted over the $edge(a,b)$.
$\lambda_{a,b}$ consists of the sum of portions of the payment rate between each pair of nodes that pass through $edge(a,b)$.
So we have:
\begin{equation} \label{rate}
  \lambda_{a,b} = \underset{s \neq t}{\sum^{}_{s,t \in V}} \frac{\sigma(s,t|edge(a,b))}{\sigma(s,t)} MRates_{st},
\end{equation}
where $\sigma(s,t)$ is the number of shortest paths from node $s$ to node $t$ and $\sigma(s,t|edge(a,b))$ is the number of shortest paths from node $s$ to node $t$ passing through $edge(a,b)$ in the directed graph G.

 \noindent \textbf{Lemma 2.} $\frac{p}{q} = \frac{\lambda(a,b)}{\lambda(b,a)}$.
\\ \\ According to Lemma 2:
\begin{equation}\label{probability_calculation}
  p = \frac{\lambda(a,b)}{\lambda(a,b) + \lambda(b,a)}
\end{equation}
Therefore we can find $p$ based on the network topology.
\\ \\ \textbf{Lemma 3.} \textit{If $\forall s,t \in V: MRates_{st} = MRates_{ts}$ then $p = 0.5$.}
\\ \\ We provide the proofs of Lemmas 2 and 3 in Appendix A.
If we assume that $MRates$ is a symmetric matrix, according to Lemma 3, $p$ is independent of $MRates$ matrix and the network topology.

\subsection {Model Analysis}

In this section we analyze the effect of channel parameters on the channel's expected lifespan and perform a financial analysis for channel lifespan.

For more realistic parameter values we used a recent snapshot of the Lightning Network taken on $Feb 2019$ as a reference point. The average payment size is considered to be $60000\,sat\footnote{satoshi}$ \cite{beres2019cryptoeconomic} and the average channel capacity is considered $2.4\,Msat\footnote{million satoshi}$ according to the snapshot.

For simplicity we use "lifespan" and "expected number of payments until channel is unbalanced", interchangeably.

\begin{figure}[h!]
  \begin{center}
  \includegraphics[width=3.0in]{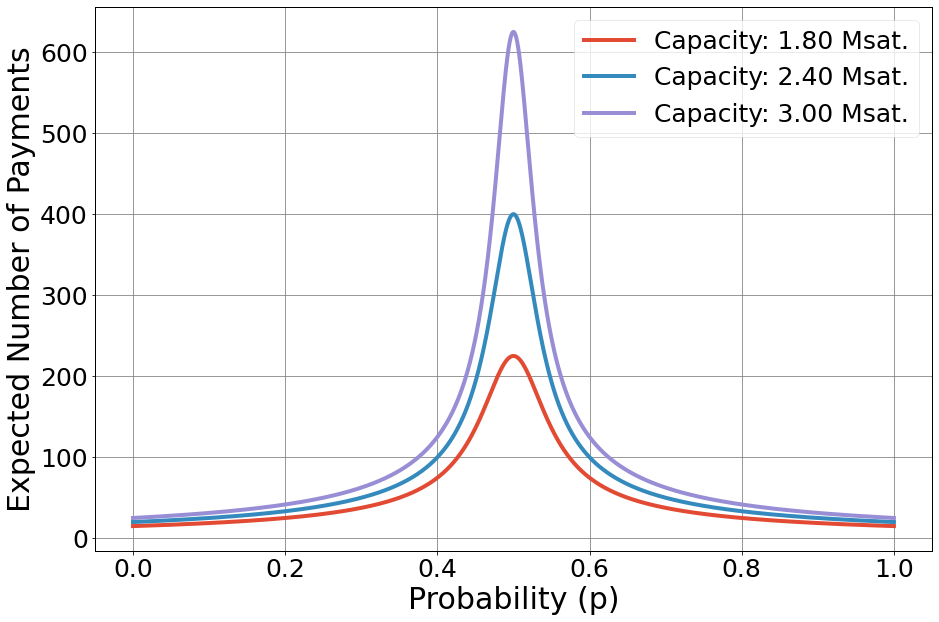}
  \caption{Effect of payment direction probability on balanced channels, according to different channel capacities.}
\label{probability_capacity}
  \end{center}
\end{figure}

\begin{figure}[h!]
  \begin{center}
  \includegraphics[width=3.0in]{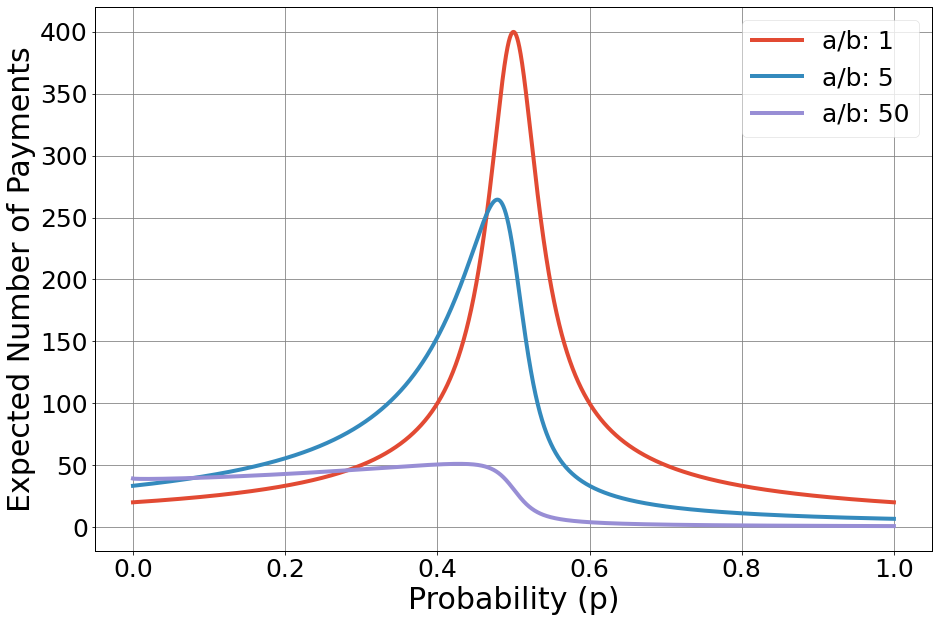}
  \caption{The effect of payment direction probability on expected number of payments, according to different initial balance ratios. The channel capacity is considered 2.4\,Msat.}
\label{probablitiy_ab}
  \end{center}
\end{figure}

We first answer the question of how sensitive is a channel's lifespan to the changes in $p$. As demonstrated in Fig. ~\ref{probability_capacity}; if the channel is initially balanced, the maximum lifespan happens on $p=\frac{1}{2}$. Also, lifespan is more sensitive to changes in $p$ when the capacity is higher. From this result we can infer that it is an important consideration for a node to make sure the channel is placed in a way that $p$ is close to $\frac{1}{2}$, otherwise the channel's lifespan is affected dramatically. A reasonable proposal for nodes who want to keep their channels active as long as possible is to charge routing fees in a way that encourages other nodes to route their payments through the node in order to achieve $p=0.5$.
Fig. \ref{probablitiy_ab} shows that if a channel is initially unbalanced, its maximum possible lifespan takes a hit. Although the maximum lifespan does not occur at exact $p=\frac{1}{2}$,  it occurs at a point close to this value. So even if a channel is somewhat unbalanced, the nodes must try to keep $p$ as close as possible to 50\%.

\begin{figure}[h!]
  \begin{center}
  \includegraphics[width=3.0in]{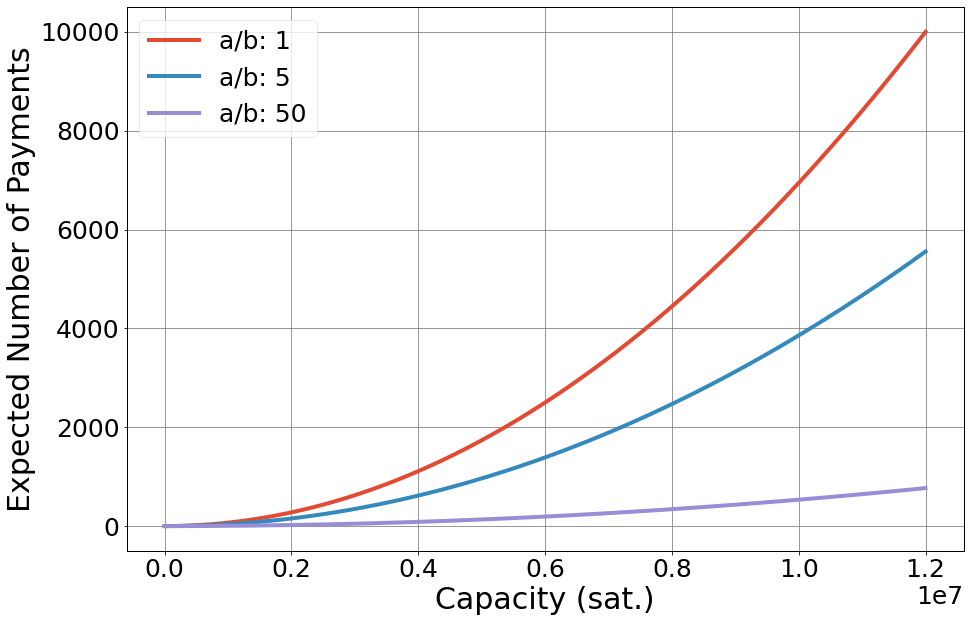}
  \caption{Effect of channel capacity on the expected number of payments, according to initial balance ratios}
\label{capacity}
  \end{center}
\end{figure}

We now answer the question of how the lifespan is affected by the channel capacity.
As Fig. \ref{capacity} suggests, the channel lifespan increases with increasing its capacity. It is noteworthy that the slope of this graph is increasing. So if a node doubles its channel capacity, the channel's lifespan will be more than doubled. Moreover, Fig. \ref{capacity} shows the effect of a channel’s initial imbalance on its lifespan.

\begin{figure}
  \begin{center}
  \includegraphics[width=3.0in]{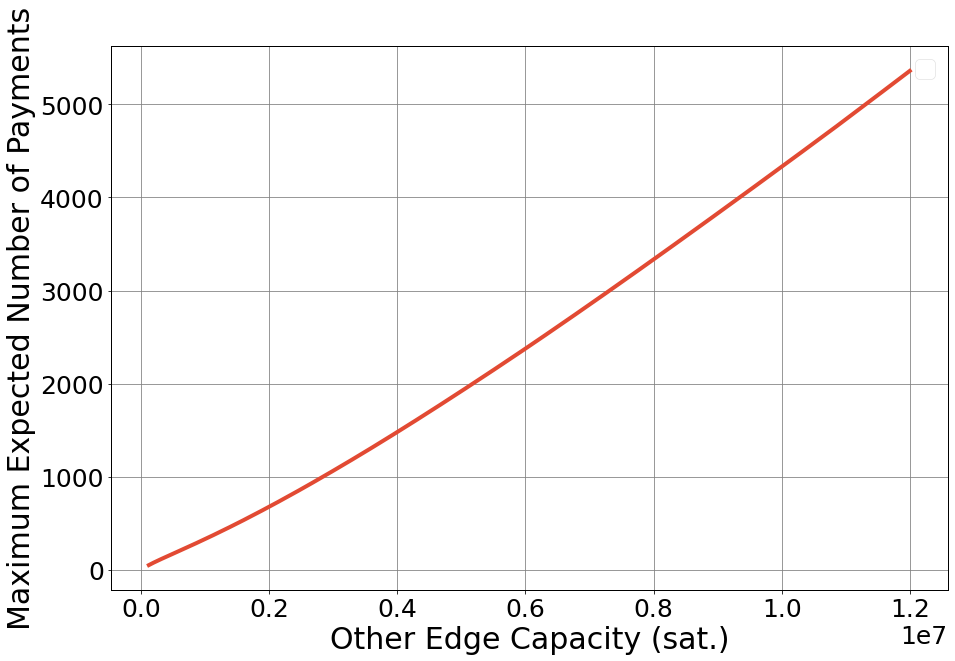}
  \caption{For a fixed a = 1.2\,Msat, the effect of channel b's capacity on the maximum possible lifespan in any p.}
\label{other_edge}
  \end{center}
\end{figure}

\begin{figure}
  \begin{center}
  \includegraphics[width=3.0in]{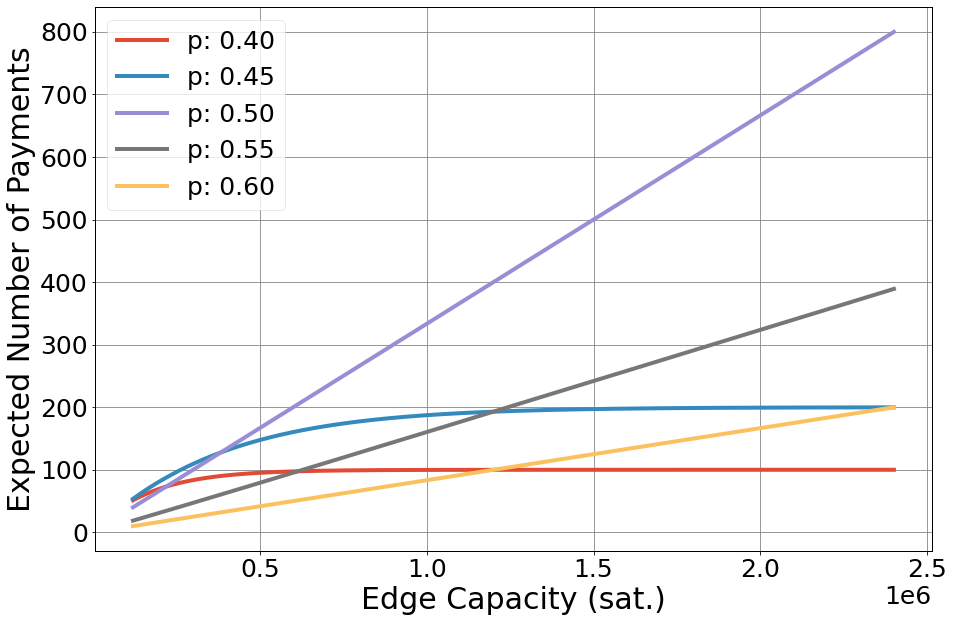}
  \caption{Having a fixed initial balance from peer node (b) analyzing the effect of our initial balance fund (a), according to different payment direction probabilities (p).}
\label{other_edge_different_p_s}
  \end{center}
\end{figure}

Usually when a node wants to create a new channel with another node in the network, the only parameter it has control over is the amount of funds it wants to put in the channel, not the funds its partner puts in the channel. This brings up the question: how will the channel's lifespan be affected with the amount the other node wants to put in the channel if our fund stays at a fixed value. Figures (\ref{other_edge}) and (\ref{other_edge_different_p_s}) illustrate this effect. Fig. ~\ref{other_edge} shows the maximum achievable lifespan considering any $p$ value and how it is affected by the fund that the other node commits to the channel. The maximum lifespan grows with the initial fund of the other edge in a linear fashion.
Fig. ~\ref{other_edge_different_p_s} illustrates the effect of our edge capacity if the peer node's capacity is fixed. Figure (\ref{other_edge_different_p_s}) shows that if $p$ is in favor of payments in the direction of our edge ($p \geq \frac{1}{2}$), the lifespan increases almost linearly; otherwise ($p < \frac{1}{2}$), the other edge becomes the bottleneck and the fund we put towards the channel will have little to no effect on the expected lifespan of the channel. If the funds we put towards the channel do not have an effect on the channel's lifespan, we have wasted cost opportunities.

\section{Implementation and Evaluation}
We provided a simulation proof of concept on a constructed Lightning Network to show the accuracy of the model discussed in Section \ref{Modeling}. In this section we describe our methodology for creating data and calculating accuracy of our model. We later analyze the results to see under which conditions the model performs better.

\subsection{Methodology}
\begin{figure}
  \begin{center}
  \includegraphics[width=3.0in]{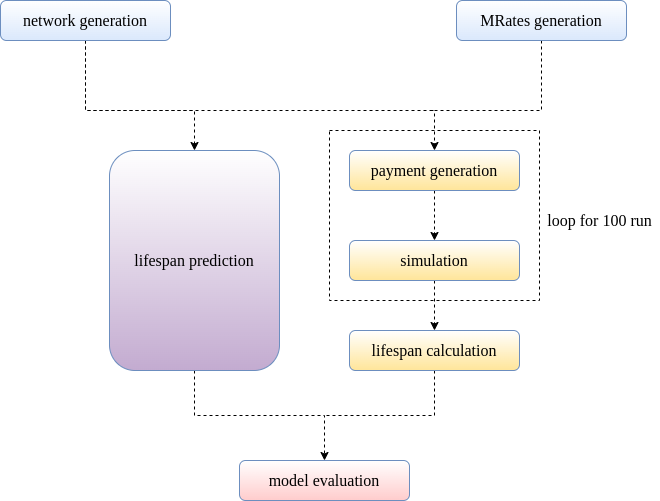}
  \caption{Model evaluation pipeline.}
\label{model_eval}
  \end{center}
\end{figure}

The testing pipeline shown in Fig. \ref{model_eval} uses the following modules:
\subsubsection{Network Generator}
For each test, a random network was generated using NetworkX's \cite{SciPyProceedings_11} $gnp\_random\_graph$ with the number of nodes being 50 and the channel existence probability being $20\%$ (245 edges on average).

\subsubsection{$Mrates$ Generator}
As discussed in previous sections the $Mrates$ matrix holds the rates in which each two nodes send payments to one another. The $Mrates$ generator takes two main parameters: SC and SK. SC determines the sparseness of the $Mrates$ matrix and SK determines the matrix skewness in relation to its main diagonal. Per each test, a new matrix is generated. In table (\ref{table:simulation_error_pred_real}) the sparse coefficient and the skew were changed to test how the model will perform in each scenario.

\subsubsection{Lifespan Predictor}
The lifespan predictor takes the network and the $Mrates$ matrix and using the model discussed in \ref{Modeling} gives the expected time for each channel to become unbalanced.

\subsubsection{Payment Generator}
Payment generator creates random payments in CLoTH simulator’s input format \cite{conoscenti2018cloth}. These payments follow the $Mrates$ generator values on average.

\subsubsection{Simulator}
We used a modified version of the CLoTH simulator \cite{conoscenti2018cloth}. We modified CLoTH such that the simulator logs the unbalancing of channels and chooses paths randomly in cases where more than one shortest path exists.

The payment generator and simulator run 100 iterations per test.

\subsubsection{Lifespan Calculator}
This module aggregates the results of 100 iterations of the previous step and calculates the average lifespan and its error. This data will be used to determine the accuracy of the model.

\subsubsection{Model Evaluation}
The error of each channel is calculated as $\frac{|real - prediction|}{real}$. 
Because some channels are positioned in a way that almost no payments pass through them, only after a long while that most channels are unbalanced, some payments pass through them, we count these channels as abnormalities and do not consider them in the error calculations. These are usually the channels that are estimated to have a very long lifespan.

The means of calculated errors are given in Table (\ref{table:simulation_error_pred_real}) considering different SC and SK values.

\begin{table}
\caption{error of prediction and real lifetime}
  \label{table:simulation_error_pred_real}
  \begin{tabularx}{\columnwidth}{X| X X X X X}
      \hline
       & & & SK & & \\
      \hline
       & & 1 & 4 & 6 & 10 \\
      \hline
       & 0.9 & 0.15 & 0.10 & 0.09 & 0.10 \\ 
      
      SC & 0.5 & 0.11 & 0.10 & 0.08 & 0.07 \\
      
      & 0.3 & 0.09 & 0.07 & 0.05 & 0.06 \\
      
       & 0 & 0.11 & 0.07 & 0.07 & 0.07 \\ 
      \hline
  \end{tabularx}
\end{table}

As we see, better results are obtained with smaller SCs (meaning a busier Lightning Network).
It is also notable that SK value has little to no effect on the model performance. This means that the model performs well in either case that $p$ is close to $0.5$ and $p$ is far from $0.5$.

\section{Lightning Network Analysis}
\label{network_analysis_section}

In this section we will provide an analysis on channel lifespans of a recent snapshot of the Lightning Network. The simulation is constituted by nodes and channels taken from a snapshot of the Lightning Network Mainnet \cite{rohrer19lnattack} on Feb 2019. 

In Section \ref{Modeling}, we proposed a model for a payment channel using a random walk and we derived a formula to predict expected channel lifespans. Moreover, the expected lifespan of a payment channel can be found if the rate of payments are known by using \ref{Expectedrateequation}. Lemma 3 shows that if we have the same rate for every pair of nodes, the probability of going to each side is equal to $0.5$.

Because payment rates and channel balances usually are not public in the Lightning Network, we have to make assumptions on the distribution, the amount of payments, and channel balances. We assume that all payment rates have the same value $r$, which means that the rates matrix ($MRates$) is symmetric. Thus according to Lemma 3, $p=\frac{1}{2}$ for every channel in the network. According to \ref{stepEquation} the expected number of payments is equal to $a \times b$, where $a = \lfloor \frac{F_{A}}{\omega}\rfloor$ and $b = \lfloor\frac{F_{B}}{\omega}\rfloor$ for a bidirectional channel between $A$ and $B$. We assume that all channels are initially balanced, meaning $a = b = \frac{C}{2\omega}$, where $C$ is the channel's capacity.

According to previous results in Section \ref{Modeling} (equations (\ref{sumRate}) and (\ref{rate})) we have:
\begingroup\makeatletter\def\f@size{8}\check@mathfonts
\begin{equation}
    \lambda_{payment}=(\underset{s \neq t}{\sum^{}_{s,t \in V}} \frac{\sigma(s,t|edge(a,b))}{\sigma(s,t)}  + \frac{\sigma(s,t|edge(b,a))}{\sigma(s,t)}) r
\end{equation}
\endgroup

We also know that
$\underset{s \neq t}{\sum^{}_{s,t \in V}} \frac{\sigma(s,t|edge(a,b))}{\sigma(s,t)}$
is equal to the edge betweenness centrality of $edge(a,b)$ ($EBC(a,b)$) in directed graph $G$ \cite{brandes2008variants}.

Because all channels are bidirectional: $\forall edge(j,i) \xrightarrow[]{} \exists edge(i,j)$, thus $\forall s,t \in V:$
\begin{equation}
  \frac{\sigma(s,t|edge(a,b))}{\sigma(s,t)} = \frac{\sigma(t,s|edge(b,a))}{\sigma(t,s)}.
\end{equation}

Assuming $G^{'}$ as an undirected graph that is derived from 
$G$ we have:
\begin{equation}
    EBC_{G}(a,b) = EBC_{G^{'}}(a,b),
\end{equation}
thus
\begin{equation}
    \lambda_{payment} = 2 \times {EBC_{G^{'}}(a,b)} \times r.
\end{equation}

If we put all results in (\ref{Expectedrateequation}), we have: 
\begin{equation}\label{timeCalc}
    E_{time} = \frac{(\frac{C}{\omega})^2}{4 \times 2 \times {EBC_{G^{'}}(a,b)} \times r}
\end{equation}

In what follows, we first calculate all payment channels’ lifespans in the LN snapshot using equation (\ref{timeCalc}). Then we focus on the relation between edge betweenness centrality and lifespan of the channels.

\subsection { Distribution of Channels Effective Lifespan }
\begin{figure}
  \begin{center}
  \includegraphics[width=3.0in]{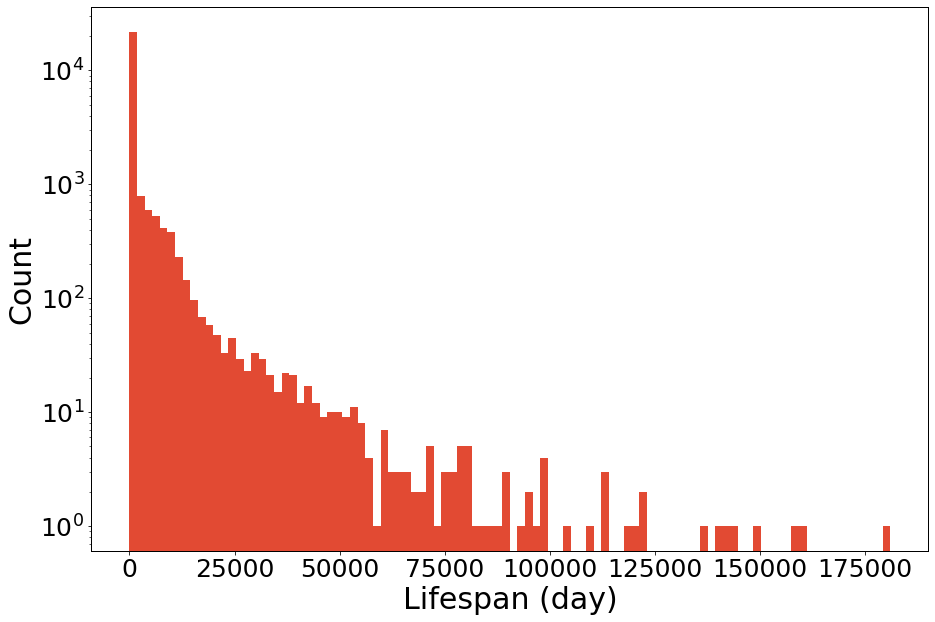}
  \caption{Histogram of expected lifespan for the LN snapshot in Feb 2019.}
\label{step_hist}
  \end{center}
\end{figure}

Equation (\ref{timeCalc}) shows that the lifespan of a channel can be calculated based on its edge betweenness centrality and initial fund. We assume that $r=0.0022$ transactions per day \cite{ratedata} and $\omega=60000\,sat$ \cite{beres2019cryptoeconomic}. The distribution of channels lifespans in our snapshot is shown in Fig. \ref{step_hist}. Much like the distribution of channel capacities that resemble the Power Law distribution, Fig.~\ref{step_hist} shows that there are a lot of channels with a low lifespan and very few channels with a very high lifespan.

According to Seres et.al. \cite{seres2020topological} the most effective channels are the channels with the highest betweenness centrality. This paper suggests that the top 14\% of the channels have the most significant effect on the network’s performance.

Table (\ref{table_simulation_parameters}) gives average, standard deviation, and median, for all channels in the network and the top 14\% central channels.

\begin{table}
        \caption{Lifespan statistics of the LN snapshot (day).}
    \begin{tabularx}{\columnwidth}{X X X}
        \hline
        & All Channels & Central Channels    \\
        \hline
        average & 1833.2 & 172.3 \\
        STD & 7086.9 & 587.2       \\
        median & 27.0 & 1.6 \\
        \hline
    \end{tabularx}
    \label{table_simulation_parameters}
\end{table}

\begin{figure}
  \begin{center}
  \includegraphics[width=3.0in]{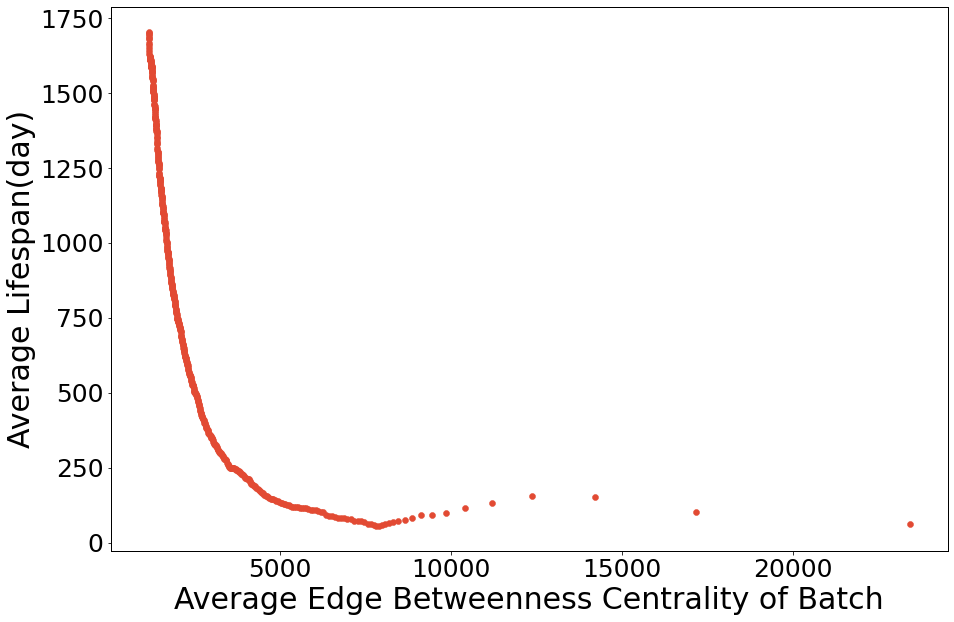}
  \caption{The relation between expected lifespan and betweenness centrality of channels in the LN snapshot.}
\label{betweenness_life}
  \end{center}
\end{figure}

\subsection {Betweenness-Lifespan Correlation}

As Seres et.al. \cite{seres2020topological} suggests, the most central channels have the most impact on the network. As Fig. ~\ref{betweenness_life} shows, more central channels have a shorter lifespan because they route more payments per unit of time. In Fig. \ref{betweenness_life} we took batches of the most central edges and calculated the average centrality and the average lifespan per batch. The result shows that in general the more central a channel is, the sooner it will get unbalanced. We see an exception to this statement in the middle of the plot, where betweenness has a positive correlation with the average lifespan. This is due to the fact that some very central edges have a large capacity so they can route more payment considering that lifespan increases with capacity quadratically.

In Section \ref{network_analysis_section} A we showed that channels with larger edge betweenness centrality values have a higher impact on the performance of the network. In this section, we have shown these central channels will have shorter lifespans. Therefore, the network success rate will decrease quickly due to unbalancing.

\section{Conclusion}
In this paper we modeled payment channel liquidity with a random walk to estimate how long it takes for a channel to become unbalanced and the effect of being unbalanced on a channel’s probability of successful routing. We also analyzed how unbalanced channels degrade the network’s performance, and the relation between a channel’s centrality and its lifespan. We showed that the network’s success probability is sensitive to the channels’ unbalancing.

We also introduced a method to estimate the lifespan of a channel in a payment network which can be used for determining a good placement in the network. We provided a proof of concept for our model and showed the results are $95\%$ accurate. 

This work shows that just allocating more funds towards a channel does not lead to having a more successful channel. The results show the channel's success in the network depends greatly on the network topology, transaction flow, and the amount of funds the peer node puts in the channel. 

We suggested the amount a node should invest in a channel to get the longest channel lifespan and maximize its return on investment. These results show that a misplaced channel can have a very short lifespan and lose up to 40\% of its efficiency, so nodes could potentially create a market based on these criteria to sell each other good connections in the network.


%

\appendices
\section{Proofs}
\subsection{\textbf{Lemma 1.} The expected number of steps to reach $+a$ or $-b$ for the first time starting from zero is 
\begin{equation} \label{stepEquation}
  E_{steps} =
  \begin{cases}
    \frac{a p^a(p^b-q^b) + b q^b(q^a-p^a)}{(p-q)(p^{a+b} - q^{a+b})} & p \neq 1/2\\
      a b & p = 1/2
  \end{cases}
\end{equation}} 
Consider $S_x$ as the expected number of steps to reach $+a$ or $-b$ for the first time starting from $x$. 
Let $p$ be the probability of going to the positive direction and 
$q$ the probability of going in the negative direction ($p+q=1$). Then we can say that
if the Random Walk starts from x, he will go to $x+1$ with probability of p and $x-1$ with probability of q.
so we can infer this recurrence equation:
$ s_x = 1 + q s_{x-1} + p s_{x+1} $
where $s_x$ is the expected number of steps until reaching the end point starting from point $x$.
For the boundary conditions we have:
$ s_a = s_{-b} = 0$ implying that the expected number of steps needed to reach $+a$ or $-b$ starting from $+a$ or $-b$ is zero.

so:

\begin{equation}\label{recurrence_equation}
  s_x = \frac{1}{p} s_{x-1} - \frac{q}{p} s_{x-2} - \frac{1}{p}
\end{equation}

The characteristic equation of (\ref{recurrence_equation}) is:
\begin{equation}
  (z^2 - \frac{1}{p} z + \frac{q}{p})(z - 1) = 0
\end{equation}

if $p = q = 1/2$ we have $\Delta = 0$ therefore $z_1 = z_2 = z_3 = 1$ so the expected number of steps needed to reach $+a$ or $-b$ starting from $x$ is:
\begin{equation}
  s_x = (a - x)(b - x)
\end{equation}

if $p \neq 1/2$ we have $\sqrt{\Delta} = |\frac{1 - 2p}{p}|$ therefore $z_1 = z_2 = 1, z_3= \frac{q}{p}$ and for the number of steps we have:
\begingroup\makeatletter\def\f@size{8}\check@mathfonts
\begin{equation}  
  s_x=\frac{a p^{a+b} + b q^{a+b}}{(2p-1)(p^{a+b}-q^{a+b})} + \frac{1}{1 - 2p} x+ \frac{(a+b)p^aq^b}{(2p-1)(q^{a+b} - p^{a+b})} (\frac{q}{p})^x
\end{equation}
\endgroup
we take $x=0$ as this gives the expected number of steps to reach $+a$ or $-b$ starting from zero. so we have:
\begin{equation}
  S_0 =
  \begin{cases}
    \frac{a p^a(p^b-q^b) + b q^b(q^a-p^a)}{(p-q)(p^{a+b} - q^{a+b})} & p \neq 1/2\\
      a b & p = 1/2
  \end{cases}
\end{equation}

\subsection{\textbf{Lemma 2.} $\frac{p}{q} = \frac{\lambda(A,B)}{\lambda(B,A)}$} In assumptions of Section \ref{Modeling} it is assumed that each node sends its payments to other nodes with a Poisson distribution. The parameter of the distribution for $edge(a,b)$ is $\lambda(a,b)$, which is the payment rate between nodes $a$ and $b$. Assume the random variable of payments from $a$ to $b$ as $X$ and the random variable of payments from $b$ to $a$ as $Y$.Thus we have:
\begin{equation}
    P(X = n) = \frac{e^{-\lambda(A,B)} (\lambda(A,B))^{n}}{n!}
\end{equation}
The total payment rate in each channel is the sum of rates of its two edges. It is known that the distribution of a random variable which is the sum of two random variables with a Poisson process is a Poisson process; the rate of this process equals the sum of rates.

When we have a payment from two Poisson distributions sending payments to the same channel; The probability for the payment to be a payment from node $a$ to node $b$ ($p$) is:
\begin{equation}
    p = P(X = 1 | X + Y = 1) = \frac{\frac{e^{-\lambda_x} (\lambda_x)^1}{1!} \times \frac{e^{-\lambda_y} (\lambda_y)^0}{0!}}{\frac{e^{-(\lambda_x + \lambda_y)} (\lambda_x + \lambda_y)^1}{1!}}
\end{equation}
Thus:
\begin{equation}
    p = \frac{\lambda_x}{\lambda_x + \lambda_y} = \frac{\lambda(a,b)}{\lambda(a,b) + \lambda(b,a)}
\end{equation}
\subsection{\textbf{Lemma 3.} If $\forall s, t \in V: MRates_{st} = MRates_{ts}$ then $p = 0.5$.}
We know from lemma 2 that: $\frac{p}{q} = \frac{\lambda(a,b)}{\lambda(b,a)}$
so we have:
\begin{equation}
  \frac{p}{q} = \frac
  {\sum^{}_{} \frac{\sigma(s,t|edge(a,b))}{\sigma(s,t)} MRates_{st}}
  {\sum^{}_{} \frac{\sigma(t,s|edge(b,a))}{\sigma(t,s)} MRates_{ts}}
\end{equation}
Because all channels are bidirectional($\forall edge(a,b): \exists edge(b,a)$) we have $\forall s,t \in V:$
\begin{equation}
  \frac{\sigma(s,t|edge(a,b))}{\sigma(s,t)} = \frac{\sigma(t,s|edge(b,a))}{\sigma(t,s)}
\end{equation}

In the other hand if we have $\forall s,t \in V: MRates_{st} = MRates_{ts}$, we can say:
\begingroup\makeatletter\def\f@size{8}\check@mathfonts
\begin{equation}
  \frac{\sigma(s,t|edge(a,b))}{\sigma(s,t)} \times MRates_{st} = \frac{\sigma(T,S|edge(b,a))}{\sigma(t,s)} \times MRates_{ts}
\end{equation}
\endgroup
then finally we have:
\begin{equation}
    \lambda(a,b) = \lambda(b,a)
\end{equation}
so 

\begin{equation}
  p = \frac{1}{2}
\end{equation}


\ifCLASSOPTIONcaptionsoff
  \newpage
\fi



%


\bibliographystyle{IEEEtran}
\bibliography{IEEEabrv,Bibliography}
%
\vfill\break
\begin{IEEEbiography}[{\includegraphics[width=1in,height=1.25in,clip,keepaspectratio]{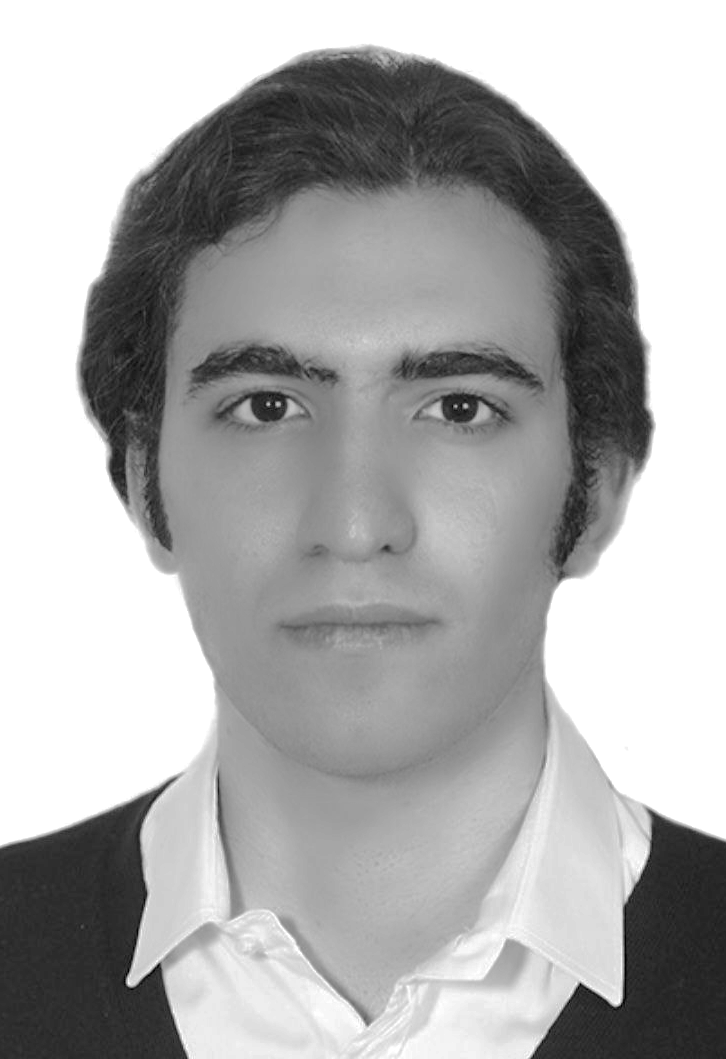}}]{Soheil Zibakhsh Shabgahi}
received his bachelor’s degree in Computer Engineering from the Department of Electrical and Computer Engineering, University of Tehran, Tehran, Iran. He is currently a research assistant at the Data lab under the supervision of professor Behnam Bahrak. His research interest are in blockchain systems, Theoretical Computer Science,distributed systems, and Machine Learning.
\end{IEEEbiography}

\begin{IEEEbiography}[{\includegraphics[width=1in,height=1.25in,clip,keepaspectratio]{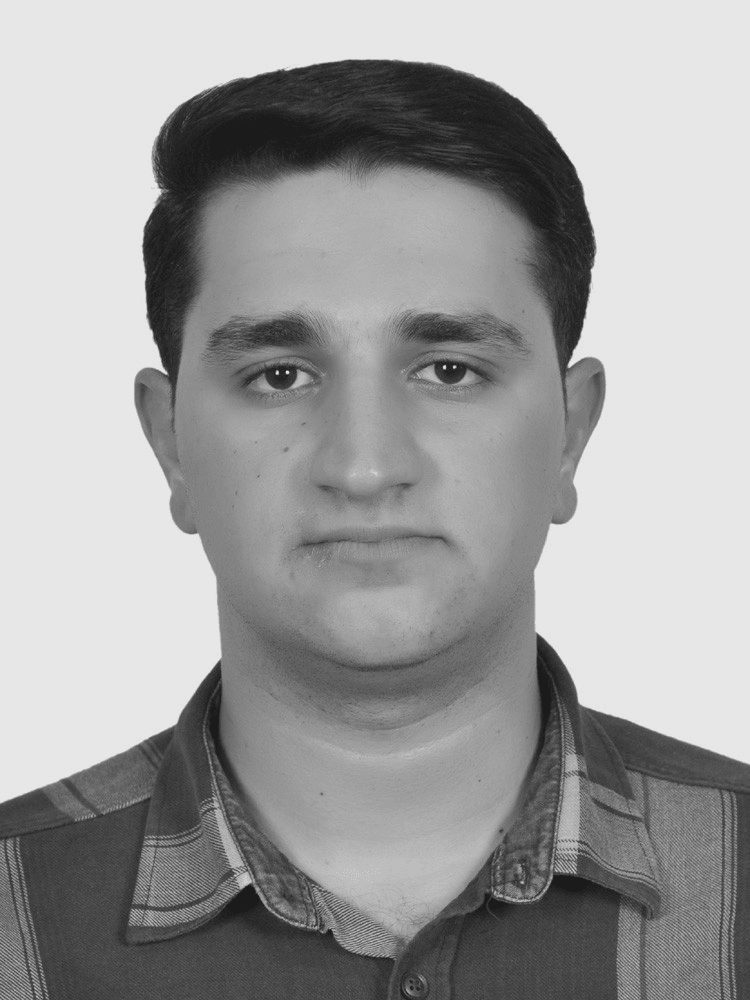}}]{Seyed Mahdi Hosseini}
is currently an undergraduate student majoring in computer engineering at the School of Electrical and Computer Engineering at the College of Engineering of the University of Tehran. His research interest consists of blockchain, system networks, and distributed systems.
\end{IEEEbiography}

\begin{IEEEbiography}[{\includegraphics[width=1in,height=1.25in,clip,keepaspectratio]{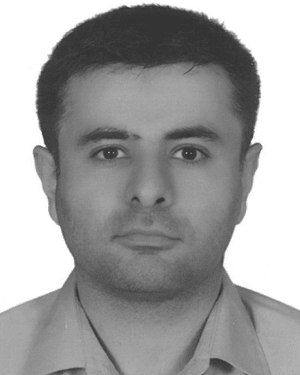}}]{Seyed Pooya Shariatpanahi}
received the B.Sc., M.Sc., and Ph.D. degrees from the Department of Electrical Engineering, Sharif University of Technology, Tehran, Iran, in 2006, 2008, and 2013, respectively. He is currently an Assistant Professor with the School of Electrical and Computer Engineering, College of Engineering, University of Tehran. Before joining the University of Tehran, he was a Researcher with the Institute for Research in Fundamental Sciences (IPM), Tehran. His research interests include information theory, network science, wireless communications, and complex systems. He was a recipient of the Gold Medal at the National Physics Olympiad in 2001.
\end{IEEEbiography}

\begin{IEEEbiography}[{\includegraphics[width=1in,height=1.25in,clip,keepaspectratio]{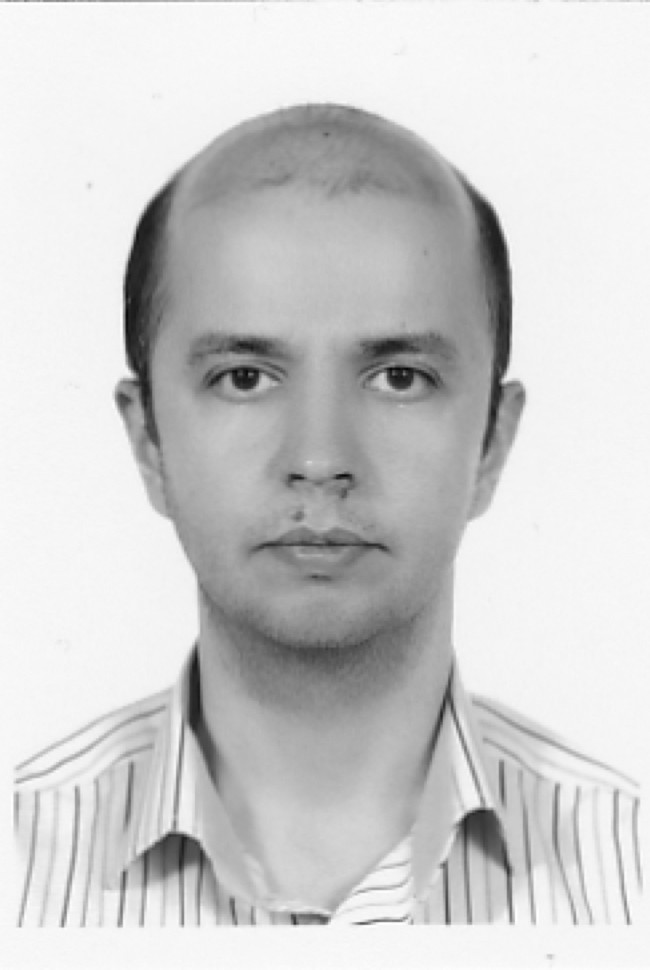}}]{Behnam Bahrak}
received his bachelor’s and master’s degrees, both in electrical engineering, from Sharif University of Technology, Tehran, Iran, in 2006 and 2008, respectively. He received the Ph.D. degree from the Bradley Department of Electrical and Computer Engineering at Virginia Tech in 2013. He is currently an Assistant Professor of Electrical and Computer Engineering at University of Tehran.
\end{IEEEbiography}
\vfill







\end{document}